\documentclass[nonacm,sigconf,authorversion,balance]{acmart}

\usepackage[utf8]{inputenc}

\usepackage{booktabs} 
\usepackage{balance}

\usepackage[most]{tcolorbox}
\usepackage{enumitem}
\newlist{questions}{enumerate}{2}
\setlist[questions,1]{label=\textbf{RQ\arabic*.},ref=\textbf{RQ\arabic*.}}
\setlist[questions,2]{label=(\alph*),ref=\thequestionsi(\alph*)}

\usepackage{xcolor}

 \newcommand\copyrighttextbottom{%
  \footnotesize Accepted version of Dominik Arne Rebro, Bruno Rossi, and Stanislav Chren. 2023. Source Code Metrics for Software Defects Prediction. In the 38th ACM/SIGAPP Symposium on Applied Computing (SAC '23), March 27-March 31, 2023, Tallinn, Estonia. ACM, New York, NY, USA. \url{https://doi.org/10.1145/3555776.3577809}}
\newcommand\copyrightnoticebottom{%
\begin{tikzpicture}[remember picture,overlay]
\node[anchor=south,yshift=24pt] at (current page.south) {\fbox{\parbox{\dimexpr1.0\textwidth-\fboxsep-\fboxrule\relax}{\copyrighttextbottom}}};
\end{tikzpicture}%
}

\begin{document}
\title{Source Code Metrics for Software Defects Prediction}

\renewcommand{\shorttitle}{Source Code Metrics for Software Defects Prediction}

\author{Dominik Arne Rebro}
\affiliation{%
  \institution{Masaryk University}
  \streetaddress{Botanicka 68a}
  \city{Brno} 
  \country{Czech Republic}}
\email{domco28@mail.muni.cz}

\author{Bruno Rossi}
\affiliation{%
  \institution{Masaryk University}
  \streetaddress{Botanicka 68a}
  \city{Brno} 
  \country{Czech Republic}}
\email{brossi@mail.muni.cz}

\author{Stanislav Chren}
\affiliation{%
  \institution{Masaryk University}
  \streetaddress{Botanicka 68a}
  \city{Brno} 
  \country{Czech Republic}}
\email{chren@mail.muni.cz}

\renewcommand{\shortauthors}{Rebro et al.}

\begin{abstract}
In current research, there are contrasting results about the applicability of software source code metrics as features for defect prediction models. 
The goal of the paper is to evaluate the adoption of software metrics in models for software defect prediction, identifying the impact of individual source code metrics. 
With an empirical study on 275 release versions of 39 Java projects mined from GitHub, we compute 12 software metrics and collect software defect information. We train and compare three defect classification models. 
The results across all projects indicate that Decision Tree (DT) and Random Forest (RF) classifiers show the best results. Among the highest-performing individual metrics are NOC, NPA, DIT, and LCOM5. While other metrics, such as CBO, do not bring significant improvements to the models.
\end{abstract}



\maketitle

\copyrightnoticebottom

\section{Introduction}

One of the crucial focuses of Software Engineering is the detection and correction of defects emerging in software systems. One approach that has been widely adopted in research is to try to predict software defects by analyzing the history of projects to identify parts of the source code that can be considered more prone to defects in the future -- utilizing the so-called software defects prediction models~\cite{Article}. 

However, while the results from existing defect prediction studies \cite{Jureczko, Article, Dambros, Deep, oomet, method_bg2, hist_bg} showcase that software metrics can be helpful in software defect prediction models, there is not an agreed view about which metrics are the best (and worse) to be used in a defect prediction model. Furthermore, the majority of the tools used in the related works are mostly deprecated or publicly unavailable. Therefore, the respective bug prediction databases created, even if public, can not be recomputed or extended with new projects (apart from Palomba et al.~\cite{Article} that provides a replication package) -- making the whole replication complicated if not unfeasible.

Using only source code metrics allows defects to be spotted very early in the development process by relying only on prior and current source code and thus resulting in potentially lower correction costs. Many related studies employ the Chidamber \& Kemerer~(CK) metric suite~\cite{Chidamber}. However, not enough analysis of which metrics from the suite are the best predictors has been conducted. The six CK suite metrics identify the size, complexity, cohesion, and coupling of source code elements and were found to correlate with defects~\cite{Chidamber}. To have a meaningful analysis of individual metrics as predictors and have a chance to compare different metric suites, we selected a metrics suite named OTHER (consisting of 5 object-oriented metrics: NPA, NPM, NLE, CBOI, CD), mainly intended to extend the CK metric suite with new uncorrelated metrics. The simple LOC metric was suggested by~\cite{LOC} to be a suitable defect predictor and will be used as a baseline for our analysis to compare the results. 
We have two major goals in this paper:
\begin{enumerate}
    \item to run an extensive study about the impact of the usage of software metrics for software defect prediction, considering individual and grouped metrics including multicollinearity issues for defects prediction. We compared LOC with CK, OTHER, and CK+OTHER software metric suites; 
    \item to identify the best (and worse) software metrics for defect prediction on open source projects selected and mined from GitHub repositories. We compared feature importance of nine metrics (LCOM5, NLE, CBO, CBOI, CD, DIT, NOC, NPA, NPM), while RFC and WMC where not considered after multi-collinearity analysis;
\end{enumerate}

The article is structured as follows. Section~\ref{sec:background} presents the area of software metrics and software defects prediction in connection with the aims of this article and details previous research related to empirical studies applying software metrics to software defect prediction. Section~\ref{sec:exp-design} presents the experimental design, with the goals, the research question, the software mining process, the metrics selection, and running the defects prediction models with the performance indicators. Section~\ref{sec:results} discusses the results in terms of the two main research questions about the impact of the usage of metrics and the most suitable metrics for defect prediction. Section~\ref{sec:threats} presents the threats to validity and Section~\ref{sec:conclusion} concludes the article.

\begin{table*}[htb]
\centering
\footnotesize
\caption{Metrics selected for defect prediction}
\label{tab:usedmet}
\begin{tabular}{|p{0.5cm}|p{7.5cm}||p{0.7cm}|p{7.5cm}|}
\hline
\textbf{Acr.} & \textbf{Description} & \textbf{Acr.} & \textbf{Description} \\ \hline
LOC & \textbf{Lines of code} -- Size metric, defined as the number of all lines of code, excluding nested, anonymous and local classes & LCOM5 & \textbf{Lack of cohesion in methods 5} -- Cohesion metric, defined as the number of functionalities of the class, with relation to the encapsulation, or single-responsibility principles. \\ \hline
WMC & \textbf{Weighted method per class} -- Complexity metric, defined as the sum of McCabe's cyclomatic complexity of the classes local methods and init blocks. & NPA & \textbf{Number of public attributes} -- Size metric, computed as the number of all public attributes, excluding attributes of nested, anonymous and local classes. \\ \hline
DIT & \textbf{Depth of inheritance tree} -- Inheritance metric, defined as the distance from the farthest ancestor in the inheritance tree. & NPM & \textbf{Number of public methods} -- Size metric, computed as the number of all public methods, excluding methods of nested, anonymous and local classes. \\ \hline
NOC & \textbf{Number of children} -- Inheritance metric, defined as the number of all directly derived classes, interfaces, enums and annotations. & NLE & \textbf{Nesting level else-if} -- Complexity metric, computed as the maximum depth of conditional, iteration and exception handling embeddedness,  considering only the fist if instruction in the if-else-if construct. \\ \hline
CBO & \textbf{Coupling between objects} -- Coupling metric, computed as the number of directly used classes, i.e. through inheritance, reference, method calls, etc. & CBOI & \textbf{Coupling between objects inverse} -- Coupling metric, computed as the inverse of CBO, i.e. the number of other classes directly using the class. \\ \hline
RFC & \textbf{Response for a class} -- Coupling metric, defined as the number of local methods and other methods directly invoked within local methods and attribute initialization. & CD & \textbf{Comment density} -- Documentation metric, computed as the ration of comment lines to total lines(i.e. comment and logical) \\ \hline
\end{tabular}
\end{table*}

\section{Background \& Related Work}
\label{sec:background}

Software defect prediction model refers to statistical models that attempt to predict potential software defects based on the test data provided~\cite{defpred}. Given the mathematical nature of software metrics, they are suitable for input for defect prediction models. Prediction models consist of independent variables (software metrics) and a dependent variable indicating defectiveness. Multiple machine learning techniques are viable for defect prediction, including regression, classification, and clustering.  
Classification uses input data referred to as a training set, which consists of objects with known class labels. The goal of a classification algorithm is to analyze the training set to develop a model that can be used to classify objects with unknown labels.
In this paper, we apply the three most popular classification algorithms used for defect prediction according to~\cite{mlstats}: \textit{Naive Bayes} (NB) \cite{MLalgos},  \textit{Decision Tree} (DT) \cite{backML} and \textit{Random Forest} (RF) \cite{RF_cite}. 

We survey the existing body of work in the area of software metric-based defect prediction in Table \ref{tab:related-work}.

\begin{table*}
\centering
\caption{Overview of metrics and prediction models in related work (full list of acronyms available in \cite{dataset2022})}
\label{tab:related-work}
\footnotesize
\begin{tabular}{|p{0.5cm}|p{0.5cm}|p{12cm}|p{3.4cm}|}
\hline
Ref & Year & Metrics & Prediction model \\ \hline
\cite{Deep} & 2020 & \textbf{Source code metrics:} 60 class-level complexity, size and object-oriented metrics & Deep neural networks \\ \hline
\cite{Article} & 2019 & \textbf{Source code metrics:}  ATFD, ATLD, CC, CDISP, CINT, CM, FanOut, LOC, LOCNAMM, MaMCL, MAXNESTING, MeMCL, NMCS, NOAM, NOLV, NOMNAMM, NPA, TCC, WOC 
\textbf{Code smell metric:} Code smell intensity & Simple Logistic, Decision Tree Majority, Naive Bayes, Logistic Regression \\ \hline
\cite{Taba} & 2013 & \textbf{Source code metrics:} LOC, MLOC, PAR, NOF, NOM, NOC, CC, DIT, LCOM, NOT, WMC 
\textbf{Process metrics:} PRE, Churn
\textbf{Antipattern metrics:} ANA, ACM, ARL, and ACPD & Various intra-system and cross-system \\ \hline
\cite{Bell} & 2011 & \textbf{Explored metrics:} log(KLOC), Prior faults, Prior changes, Prior changed, Prior developers, Prior lines added/deleted/modified & Binomial regression \\ \hline
\cite{Jureczko} & 2010 & \textbf{Source code metrics:} WMC, DIT, NOC, RFC, LCOM, CBO, LCOM3, NPM, DAM, MOA, MFA, CAM, IC, CBM, AMC, Ca, Ce, CC, LOC & Kohonen's neural network \\ \hline
\cite{Dambros} & 2010 & \textbf{Source code metrics:} WMC, DIT, NOC, RFC, LCOM, CBO,LOC, FanIn, FanOut, NOA, NPA, NOPRA, NOAI, NOM, NPM, NOPRM, NOMI
\textbf{Change metrics:}  EDHCM, LDHCM, LGDHCM & Generalized linear regression \\ \hline
\cite{Hassan} & 2009 & \textbf{Change complexity metrics:} BCC, ECC, HCM & Statistical linear regression \\ \hline
\cite{oomet} & 2001 & \textbf{Inheritance metrics: }DIT, NOC
\textbf{Coupling metrics:} ACAIC, ACMIC, DCAEC, DCMEC, OCAIC, OCAEC, OCMIC, OCMEC & Calibrated logistic regression \\ \hline
\end{tabular}
\end{table*}

Some of the authors \cite{Jureczko, Deep, oomet} focus on defect prediction directly based on the source code metrics. In particular, Jureczko et al.~\cite{Jureczko} conducted a study regarding the clustering of software projects about software defect prediction using 19 class-level source code metrics on 38 proprietary, open source, and academic projects. Emam et al.~\cite{oomet} focused on object-oriented source code metrics. Their study followed hypotheses stated in~\cite{qfactors}, that structural class properties affect the cognitive complexity (i.e., the mental burden of individuals working with the component), which in turn impacts external attributes such as fault-proneness or maintainability.

Other authors, such as Palomba et al. \cite{Article}, and Taba et al. \cite{Taba}, introduced new metrics related to the presence of code smells or antipatterns that might indicate faulty classes. The new metrics were based on the existing source code metrics. Additionally, Taba et al.~\cite{Taba} incorporated process metrics in the prediction. Both studies concluded that the presence of code smells is a relevant predictor in models. 

The evaluation of code changes has frequently complemented the prediction based on source code metrics. Hassan et al.~\cite{Hassan} proposed multiple metrics based on code change complexity as a superior predictor to past faults or modifications. Bell et al.~\cite{Bell} validated several code change metrics proposed by Moser et al.~\cite{Moser}, with the rationale being that frequently changed code elements are more likely to contain defects. They investigated to what degree faults can be predicted using code churn alone and in combination with other characteristics. The studies demonstrated that metrics based on prior code changes performed well in bug prediction. One of the most extensive studies combining the analysis of single-version and multiple-version projects was conducted by D'Ambros et al.~\cite{Dambros}. Their work resulted in a dataset consisting of source code metrics,  previous defects, change metrics, code churn, and entropy of changes. This makes it suitable for benchmarking the performance of new prediction models. On the other hand, the granularity of their analysis was at the file level rather than class level, which prevents analysis of cases when one file contains multiple nested classes.

While different studies often used different metrics, several overlaps can be observed. These include variants of the LOC metric, CK  metric suite~\cite{Chidamber}: WMC, DIT, NOC, RFC, LCOM, CBO. These metrics seem suitable for defining a baseline, against which novel metrics can be compared, and have been frequently used in researched literature. 

Moreover, various prediction results scoring metrics are used across the studies. For classification models, the most common being precision, recall, F-measure, and AUC-ROC \cite{Deep,Article,oomet}, while for regression models, metrics such as mean squared error, $R^{2}$ and Spearman's correlation coefficient are used \cite{Bell,Jureczko,Dambros,Hassan}. Some papers focused on the comparison between metrics \cite{Deep,Taba,Jureczko}, while others concentrate on pure prediction model performance \cite{Article,Bell,Dambros,Hassan,oomet}. This inconsistency makes the comparison of results across all related works problematic. However, in general, the results indicate that models using the Random Forest algorithm tend to achieve the best classification performance as observed in~\cite{Deep, GH2, tradproc}, with the Decision Tree and Simple Logistic algorithms lagging slightly behind \cite{Jureczko}.

While investigating the related work, we noticed that some studies share the datasets publicly~\cite{Deep,Article,Taba,Jureczko,Dambros}. However, most reviewed studies only provide the experiment results. The entire experimental package is publicly available only for Palomba et al.~\cite{Article}. Thus most of the studies could be challenging to reproduce, even if the analysis was done on open source projects.

\section{Experimental Design}
\label{sec:exp-design}

We have two main Research Questions (RQs) in this research.

\textbf{RQ1.} \textit{What is the impact of code metrics on software defect prediction?} The rationale of RQ1 is about evaluating how the inclusion of software metrics (and their specific groupings) can improve the software defect prediction results.

\textbf{RQ2.} \textit{Which metrics are the most suitable predictors?} The rationale of RQ2 is about identifying which source-code level software metrics considered in this article are the best (and worse) predictors in the context of software defect prediction.

\textbf{Data Source.} To build the dataset for this study, we opted for mining the GitHub repositories. With open source projects, different metrics and prediction models can be tested on the same source code, allowing for better comparison of the results.

\textbf{Project selection.} In total 39 suitable Java projects, consisting of 275 versions, were selected for our dataset. The details of project selection process, together with all the projects are available on Figshare~\cite{dataset2022} -- including statistics regarding the repository size, development team size, and bug-tracking information. More details are available also in~\cite{Rebro2022thesis}.

\textbf{Metrics selection.} For the defect prediction analysis, the following 12 metrics were selected: (1) baseline metric LOC indicates the source code size for each class; (2) The CK~\cite{Chidamber} metric suite consists of six object-oriented metrics WMC, DIT, NOC, RFC, LCOM5 (a modified version of the original LCOM metric), CBO; (3) The OTHER metric suite is consisting of five additional object-oriented metrics NPA, NPM, NLE, CBOI, CD. 
    
        \begin{itemize}
            \item Baseline metric LOC indicates the source code size for each class 
            and, therefore, the probability of defect occurrence since larger classes are more likely to contain faults than smaller ones. 
            
            Additionally, it has been suggested by~\cite{LOC} to be a suitable predictor of defects. 
            However, because of its high correlation with many other product metrics, it has been used as a stand-alone baseline metric in many other related works~\cite{Dambros}.
            
            \item The CK~\cite{Chidamber} metric suite consists of six object-oriented metrics WMC, DIT, NOC, RFC, LCOM5 (a modified version of the original LCOM metric), CBO. 
            These metrics were validated by~\cite{Basili} on several C++ projects, showing their usefulness as a quality indicator. 
            The CK suite has been extensively used for defect prediction. However, the comparison of the impact of its individual metrics on defect prediction has been mostly disregarded in studies with more extensive datasets, making it suitable for our \textbf{RQ2} feature importance analysis.
            
            \item The OTHER metric suite is consisting of five additional object-oriented metrics NPA, NPM, NLE, CBOI, CD. 
            These metrics were selected because they could potentially include new information in the prediction model due to some conceptual differences with the CK metrics. 
            This also means that most projects' correlation between the OTHER and CK metrics is low. 
        \end{itemize}

\begin{figure*}
        \centering
        \includegraphics[scale=0.5]{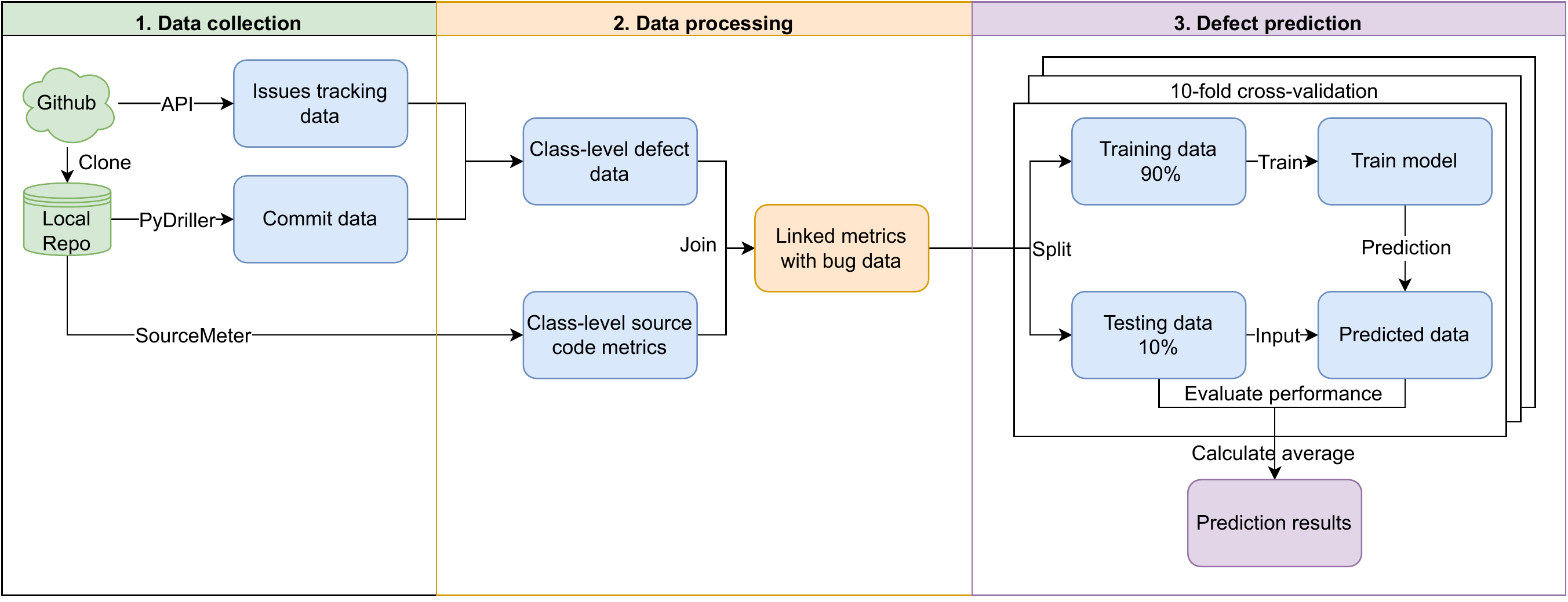}
        \caption{Overview of the experimental process}
        \label{fig:process}
    \end{figure*}

\textbf{Data collection.} For implementing the whole data collection, processing and defect prediction process, we used Python libraries PyGithub, PyDriller and scikit-learn. An overview of the whole process is presented in Fig. \ref{fig:process}.
We employed an approach for linking fault information to system versions by \cite{Bell, Hassan}. This approach allows for accumulating numerous bugs into a single version, the source code of which can be analyzed to calculate software metrics. 
Due to the object-oriented nature of the projects considered and class-level metrics being found more effective than file-level metrics, the dataset is created considering only class-level metrics by using the static code analysis tool SourceMeter.
By parsing diff files available for all files modified in a commit, we can learn which specific lines of a file were modified. With discovered source code positions of all classes, we can match the modified lines to specific classes and mark them as defective. 

\textbf{Data processing.} Defect prediction models can suffer from multicollinearity, where two or more metrics of the model are highly correlated. Multicollinearity distorts feature importance analysis related to \textbf{RQ2}. In our case, feature importance values for each metric were computed using a 10-fold cross-validated permutation feature importance~\cite{rf} model, defined as the decrease in the model score when a single feature value is randomly shuffled~\cite{rf}. 
    
To address this issue, the  Variation Inflation Factor (VIF)~\cite{thompson2017extracting} is usually used in related research~\cite{Taba, Article}. It is a measure of multicollinearity defined as the ratio of overall model variance to the variance of a model consisting of a single independent variable~(metric). 
By calculating the VIF value for each metric, it is possible to detect variables suffering from multicollinearity. As suggested in~\cite{Taba} the threshold of VIF was set to 2.5 for further investigation, while VIF values above 10 would indicate serious multicollinearity.   
While~\cite{Basili} found a correlation between RFC and CBO, in~\cite{Article} the pair of metrics RFC and WMC was found to be highly correlated, which led to the removal of RFC from further analysis. Even after disregarding RFC, the WMC metric consistently achieved VIF values above 5. Therefore we decided to remove both metrics RFC and WMC from the feature importance analysis relating \textbf{RQ2}.
In our study, the VIF values of all metrics were computed for each project.  
In a couple of outlying cases, some metrics achieved VIF close to 10, indicating significant multicollinearity. We concluded that there will probably always be an outlier project for any metric combination that suffers from high multicollinearity. In such cases, we simply disregarded these outlier projects from \textbf{RQ2} feature importance analysis.
    
Generally, defective parts of the system represent a small minority, meaning the dataset can be skewed toward classes without bugs. This means that it might prove beneficial to first use under-sampling to eliminate the class imbalance, as this can improve the accuracy of prediction models. However, too much under-sampling can lead to loss of information and thus, the underfitting of the model, which can cause poor bug prediction results. For this reason, the class imbalance was only addressed to the extent of deleting duplicate rows, which were almost exclusively of the majority ($not$ $defective$) class.

\textbf{Defect Prediction.} We treat the prediction problem as a binary classification problem, assigning each class into the $defective$ or $not$ $defective$ category. 
The processed dataset is divided into a training and testing subsets. The training set is used to fit (train) the prediction model, meaning the algorithm is learning how to predict the dependent variable. The testing set predicts dependent variable values and validates them based on scoring metrics such as precision, recall, F-measure, or AUC-ROC.  
For a better estimate of our models' performance, 10-fold cross-validation was employed by randomly splitting the data into 10 stratified folds, with 90\% of the data used for training and 10\% for testing. 

\section{Results}
\label{sec:results}

\subsection{Impact of code metrics (RQ1)}

Regarding the performance of our defect prediction models, we evaluated the results separately for each metric suite to compare their effectiveness. For each scoring metric, we provide boxplots describing the distribution of results of each model. F-measure is computed taking into account the defective (minority) class as we believe this is the most useful scenario for defects prediction. AUC-ROC is instead computed using weighted averaging per class.

 \begin{figure}[htb]
        \begin{center} \includegraphics[scale=0.65]{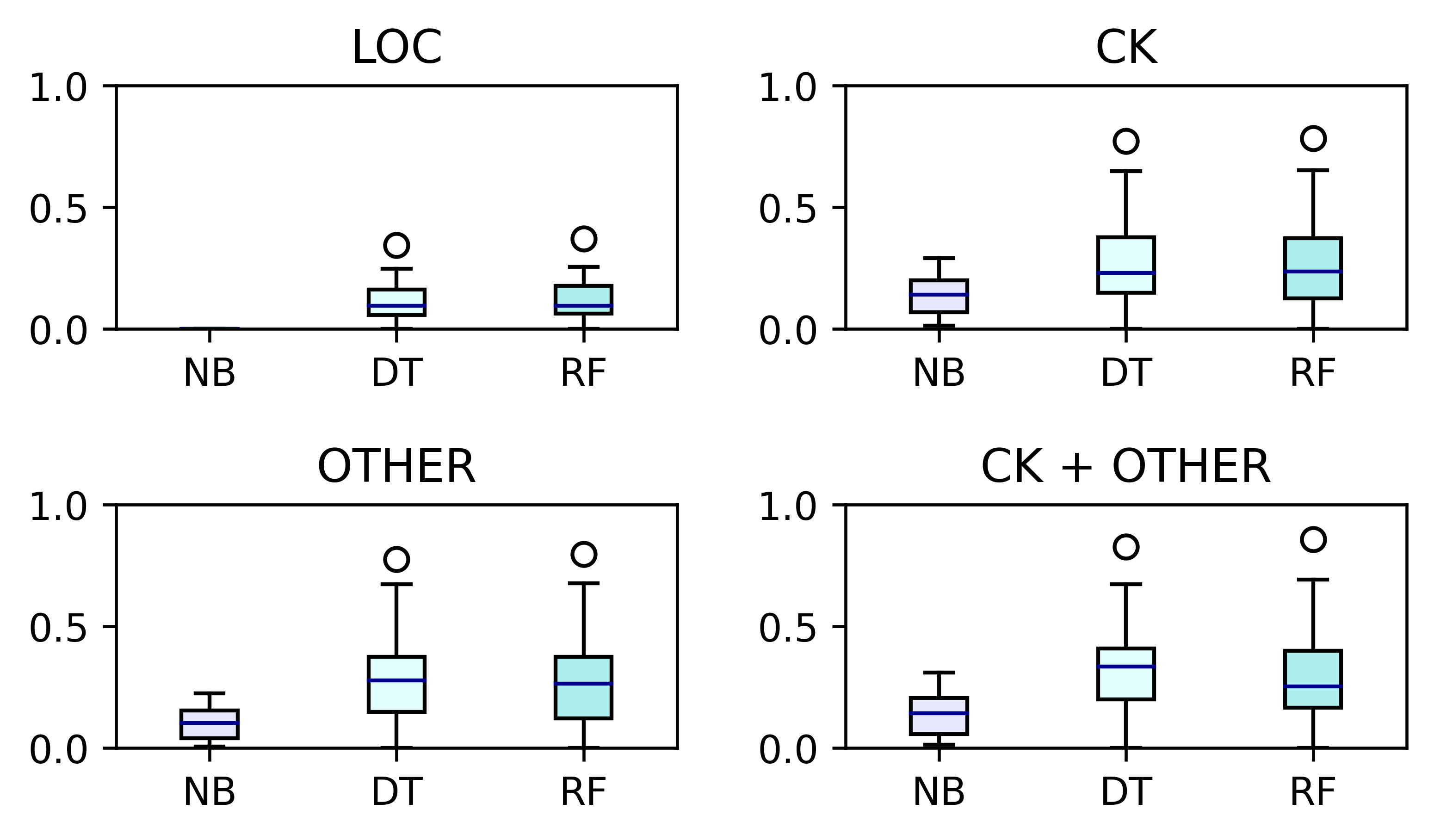}
        \end{center}
        \caption{F-measure distribution by metric suites}
        \label{fig:fsc}
\end{figure}

Looking at the F-measure distribution in Fig. \ref{fig:fsc}, we can observe that in every metrics suite models, DT and RF achieve very similar results, while NB is performing significantly worse overall. This confirms previous research suggesting the suitability of DT and RF models for defect prediction~\cite{Deep, GH2}. By comparing the interquartile range (dispersion) and median (location) of boxplots of models DT and RF, we can infer that using our defect dataset, the DT is slightly superior across all metric suites to RF in terms of F-measure -- however, such differences are not statistically significant. Running Mann-Whitney U Test comparing the mean ranks, the differences are statistically significant for all the cases of NB vs DT and RF -- while differences between DT and RF are not statistically significant. Running Kruskal-Wallis One-Way ANOVA on all the groups indicates statistically  significant differences for all the four cases (LOC, CK, OTHER, CK+OTHER).
    
Comparing the F-measure between metric suites, as expected, LOC achieved the worst and least dispersed results. The differences between the distribution of results for suites CK, OTHER, and CK+OTHER are much more subtle. Overall, we found out similar results.
While in~\cite{Article} achieved F-measures are mostly between 50\% and 80\%, in our case the results are much lower due to the fact that F-measure was computed on the defective (minority) class -- this explains F-measure lower than 50\% in some cases.

    \begin{figure}[htb]
        
        \begin{center}
            \includegraphics[scale=0.65]{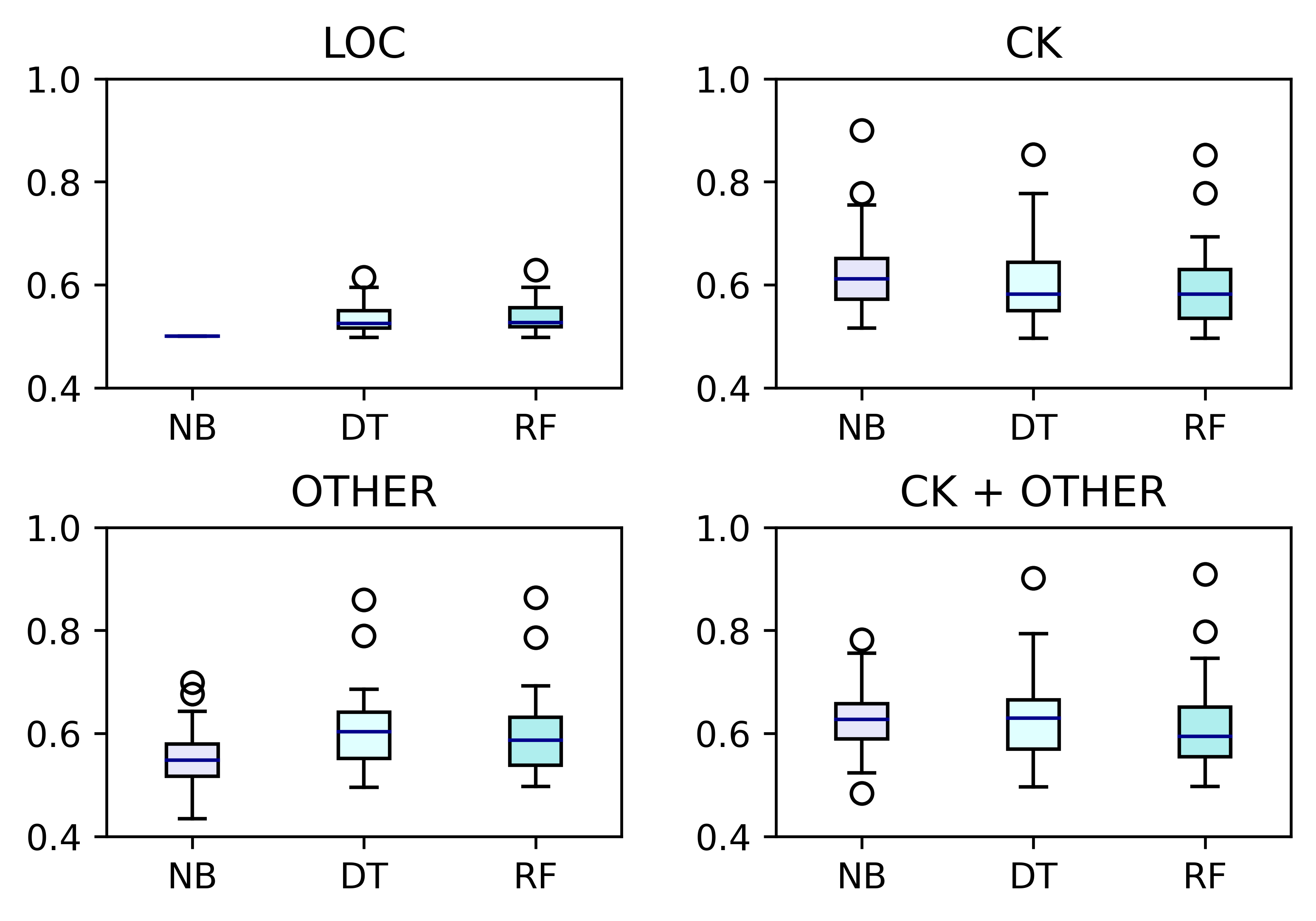}
        \end{center}
        \caption{AUC-ROC distribution by metric suites}
        \label{fig:ar}
        \end{figure}
    
The AUC-ROC scores found in Fig. \ref{fig:ar}, reports slightly different results than the F-measure. In the case of LOC and OTHER suites the NB model is still underperforming. However, in suites CK and CK+OTHER, the NB model demonstrates slightly superior performance. However, it is essential to note that the results for NB models are marginally more unstable, also achieving AUC-ROC marginally under 50\%, which brings concerns about the suitability of NB for defect prediction.  Running Mann-Whitney U Test comparing the mean ranks, the differences are statistically significant for LOC, CK, OTHER in the cases of NB vs DT and RF but not significant in CK+OTHER case. Differences between DT and RF are not statistically significant in all cases. Running Kruskal-Wallis One-Way ANOVA on all the groups indicates statistically  significant differences on two cases out of four (LOC, OTHER).
    
In terms of comparing AUC-ROC between metric suites, we can conclude that LOC is performing the worst, while other metrics suites CK, OTHER, CK+OTHER are similarly effective for defects prediction.
    
Overall, the achieved AUC-ROC scores are mostly between 50\% and 75\%, which is marginally worse than the results in some of the related research~\cite{Deep, Article}, which achieved AUC-ROC scores ranging from around 60\% to 80\%.

\begin{tcolorbox}[colback=gray!5!white,colframe=gray!75!black,title=RQ1 Summary]
       For defects prediction, the best results were achieved by DT and RF models.  Using CK, OTHER, and CK+OTHER metric suites provides similar results, but provides improvements over the standard LOCs metric.
\end{tcolorbox}

\subsection{Most suitable defect predictors (RQ2)}

\begin{figure}[htb]
        \begin{center}
            \includegraphics[scale=0.67]{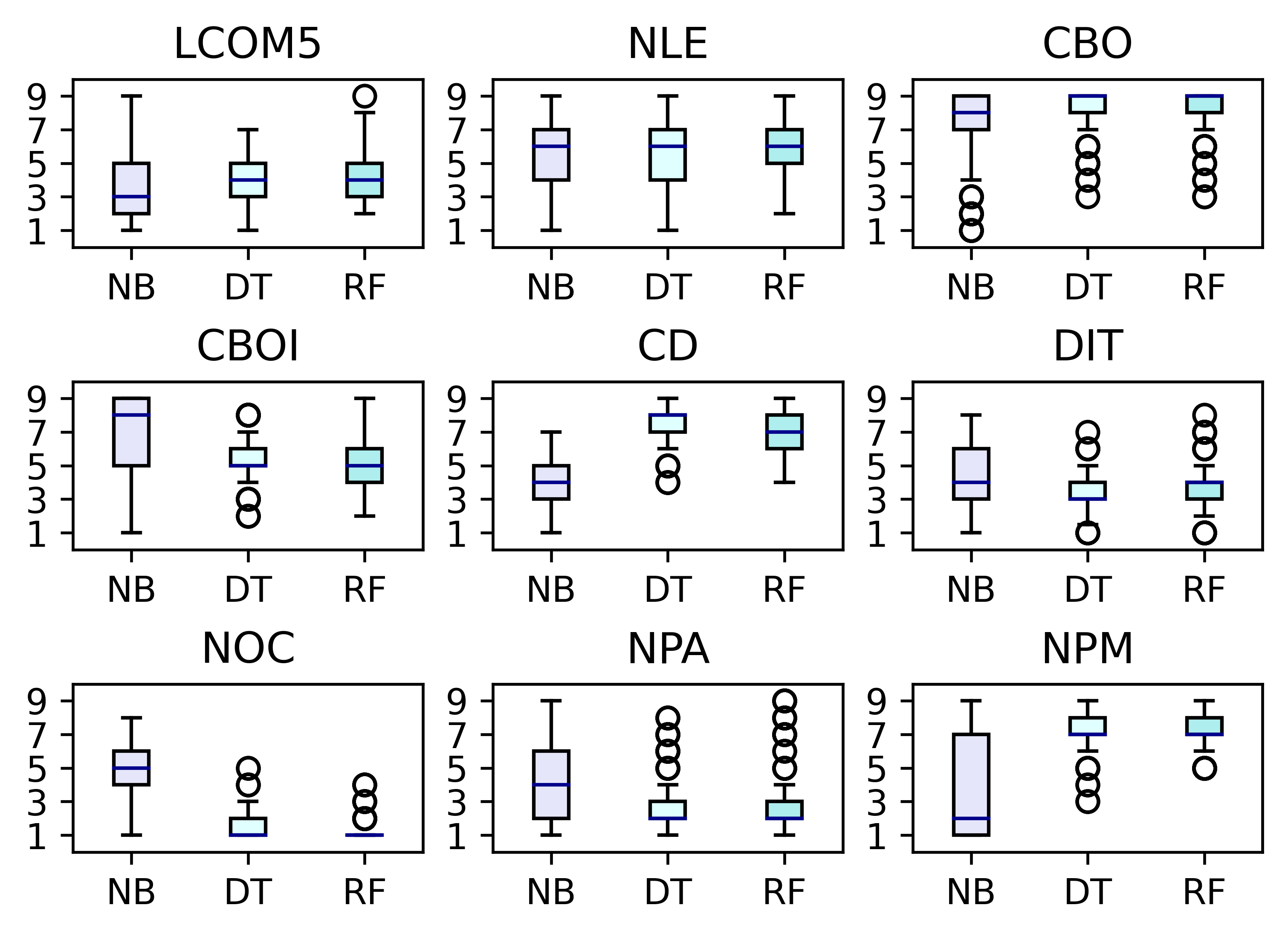}
        \end{center}
         \caption{Rankings for all metrics (the lower the better)}
        \label{fig:box-rankings}
        \end{figure}

To identify which metrics had the most positive influence on the defect prediction, we inspect each metric's distributions of feature importance ranks for all suitable projects. The boxplots of metric ranks for every project in the aggregated form are shown in Fig.~\ref{fig:box-rankings}. The individual boxplots for each project are available on Figshare~\cite{dataset2022}.

While carefully analyzing the distributions of feature importance ranks for each metric across all projects, we observed multiple trends. NOC is by far the best-performing metric, being ranked first across the vast majority of instances. Although less stable, metrics NPA, DIT, and LCOM5  tend to rank between 2\textsuperscript{nd} and 4\textsuperscript{th}. These metrics, together with NOC, consistently have a significant positive influence on the defect prediction models.

Another stable trend observed is that CBO performed consistently worst, ranked last across most projects. With slightly more variability metrics, CD and NPM consistently achieved ranks 6\textsuperscript{th} to 8\textsuperscript{th}. This indicated that these metrics lack predictive power and contribute less to the overall defect prediction results. On the other hand, it could still be because of prevalent multicollinearity, which we addressed when processing the data.

\begin{tcolorbox}[colback=gray!5!white,colframe=gray!75!black,title=RQ2 Summary]
          The best-performing metric across all projects is NOC, followed by NPA, DIT, and LCOM5. The worst-performing metric across all projects was CBO, followed by CD and NPM.
\end{tcolorbox}

\section{Threats to Validity}
\label{sec:threats}

\textbf{Construct Validity} concerns the relationship between theory and observation~\cite{consthr}. The first threat relates to how defects are linked to project versions and subsequently to classes: we used a heuristic that relies on commit messages to identify defect-fixing commits. For this reason, any defect not mentioned in a commit message will not be identified. However, the adopted heuristic represents state-of-the-art in identifying defects in versioning system files and was adopted in previous studies~\cite{Dambros, buglink}. 
    
Another construct validity threat concerns the way defects are linked to software releases. Firstly, defects are assigned to releases based on the time of the creation of the related issue. While it would be desirable to link software defects to the version in which they are introduced, this information is not available from issue tracking systems. Secondly, the way defects are aggregated to versions is distributed in a 6-months interval -- otherwise, not enough defect data would be present for each version. However, this is a known limitation, as it has been used in previous software defect prediction research~\cite{GH2, Article}.
    
\textbf{Internal Validity} concerns the selection of tools and methods of analysis~\cite{extthr}. Here we can report a threat regarding analyzing the features' importance of individual metrics. As discussed in Section~\ref{sec:exp-design}, the results of this analysis are impacted by the multicollinearity between metrics. We addressed this threat utilizing the VIF measure, removing the most problematic metrics from the feature importance analysis: the remaining metrics achieved \textasciitilde~2.5 VIF, which was considered an appropriate threshold in previous studies~\cite{Taba}.

\textbf{External Validity} concerns the generalizability of the results outside the context of the study~\cite{extthr}. The dataset built by mining software repositories comprised 275 release versions from 39 open source Java projects. To minimize the external validity threat, we sampled the projects from various application domains, considering different characteristics such as code size, number of classes, and rate of defects. According to empirical software engineering research, we can define this selection strategy as \textit{purposive sampling}~\cite{baltes2022sampling} so that projects are selected attempting to maximize the heterogeneity of the descriptive statistics. Nevertheless, the results only represent the context of Java applications, as we did not consider other software development ecosystems.

\section{Conclusion}
\label{sec:conclusion}

In this paper, we looked at the impact of software source code metrics on defect prediction models. We evaluated the performance of three classifiers in terms of several accuracy metrics and then the rankings of the metrics in terms of contribution to the prediction results.
We mined software projects from GitHub data, building a dataset of 39 Java projects (275 release versions in 6-months ranges). The creation process integrated software defects information and project source code used to calculate 12 class-level software metrics and identify the classes containing software defects. 
Overall, we can report that DT and RF had the best results compared to NB and that using software metrics for defect prediction improves the performance of the classifiers. Related to the metrics with the most positive influence on defects prediction, we found that metrics NOC, NPA, DIT, and LCOM5 were consistently ranked as the best across all projects. Conversely, the metrics CBO, CD, and NPM were classified as the worst predictors.

\begin{acks}
The work was supported by ERDF "CyberSecurity, CyberCrime and Critical Information Infrastructures Center of Excellence" (No. CZ.02.1.01/0.0/0.0/16\_019/0000822).

\end{acks}

\balance

\bibliographystyle{ACM-Reference-Format}
\bibliography{refs} 

\end{document}